\documentstyle[epsfig,12pt]{article}

\newcommand\Jou[4]{{#1} #2 (#4) #3}

\newcommand\NPA{{Nucl. Phys.} A}
\newcommand\NPB{{Nucl. Phys.} B}
\newcommand\PLB{{Phys. Lett.}  B}
\newcommand\PR{{Phys. Rep.}}
\newcommand\PRL{Phys. Rev. Lett.}

\newcommand\PRD{{Phys. Rev.} D}

\newcommand\ZPA{{Z. Phys.} A}
\newcommand\ZPC{{Z. Phys.} C}
\newcommand\EPJC{{Eur. Phys. J.} C}

\newfam\BMath
\font\BMathL=cmmib10 
\font\BMathl=cmmib7
\font\BMathm=cmmib5
\textfont\BMath=\BMathL \scriptfont\BMath=\BMathl
\scriptscriptfont\BMath=\BMathm

\newcommand\B{{\fam\BMath b}}

\newcommand\K{{\fam\BMath k}}

\renewcommand\a{\alpha}

\renewcommand\c{\chi}
\renewcommand\d{\delta}

\newcommand\e{\epsilon}
\newcommand\f{\phi}
\newcommand\g{\gamma}
\renewcommand\j{\Psi}
\newcommand\J{\psi}
\renewcommand\l{\lambda}

\newcommand\p{\pi}

\newcommand\m{\mu}
\newcommand\n{\nu}

\newcommand\r{\rho}
\newcommand\s{\sigma}

\newcommand\cc{{\cal C}}
\newcommand\cb{{\cal B}}

\newcommand\cm{{\cal M}}
\newcommand\co{{\cal O}}
\newcommand\cp{{\cal P}}

\newcommand\dd{\mbox{d}}

\renewcommand\exp{\mbox{\rm exp}}

\newcommand\ra{\rightarrow}

\newcommand\lra{\longrightarrow}
\newcommand\llra{\longleftrightarrow}

\newcommand\srm[1]{\mbox{\sevenrm #1}}

\newcommand\dxb{[\dd x]}
\newcommand\dyb{[\dd y]}
\newcommand\dkb{[\dd^2 \Kp]}
\newcommand\lqcd{\Lambda_{\rm {QCD} }}

\newcommand\intksno[1]{\frac{\dd^2 \K_{\perp\,#1}}{(2\pi)^2}}
\newcommand\intbs[1]{\frac{\dd^2 \B_#1}{(4\pi)}}

\newcommand\intbds[1]{\frac{\dd^2 \B'_#1}{(4\pi)}}
\newcommand\Kp{{\K_\perp}}

\newcommand\Kpn[1]{\K_{\perp \,#1}}
\newcommand\mc{(2 m_c)}
\newcommand\mcj{M_{\c_J}}
\newcommand\mcjo{M_{\c_1}}
\newcommand\mcjt{M_{\c_2}}
\newcommand\mjp{M_{J/\J}}
\newcommand\ot{\otimes}

\newcommand\be{\begin{equation}}
\newcommand\ee{\end{equation}}
\newcommand\bea{\begin{eqnarray}}
\newcommand\eea{\end{eqnarray}}
\newcommand\ba{\begin{array}}
\newcommand\ea{\end{array}}
\newcommand\eref[1]{Eq.~(\ref{#1})}
\newcommand\ederef[2]{Eqs.~(\ref{#1}) and (\ref{#2})}
\newcommand\etrref[3]{Eqs.~(\ref{#1}), (\ref{#2}) and (\ref{#3})}
\newcommand\equref[4]{Eqs.~(\ref{#1}), (\ref{#2}), (\ref{#3}) and (\ref{#4})}

\newcommand\fref[1]{Fig.~\ref{#1}}
\newcommand\tref[1]{Table~\ref{#1}}
\newcommand\bfi{\begin{figure}}
\newcommand\efi{\end{figure}}
\newcommand\bpi[1]{\begin{picture}#1}
\newcommand\epi{\end{picture}}

\newcommand{\ncom}{\newcommand}
\ncom{\lan}{\langle}
\ncom{\ran}{\rangle}
\ncom\fx{\!\!\!\!}

\font\lbigbf=cmbx12 scaled 1500
\font\sevenrm=cmr7

\begin{document}


\begin{centering}
{{\lbigbf Colour Octet Contribution in Exclusive P-Wave}

\vspace{8pt}
{\lbigbf Charmonium Decay into Proton-Antiproton}}

\vspace{0.25cm}

{S.M.H. Wong}

\ \\
{\em School of Physics and Astronomy, University of Minnesota, Minneapolis, \\
Minnesota 55455, U.S.A.\footnote{present address}
\footnote{email: swong@nucth1.hep.umn.edu}} \\
\ \\ 
{\em Fachbereich Physik, Universit\"at Wuppertal,
D-42097 Wuppertal, Germany} \\
\ \\
{\em Nuclear and Particle Physics Section, University of Athens,} \\ 
{\em Panepistimiopolis, GR-15771 Athens, Greece}  \\
{\em and} \\
{\em Institute of Accelerating Systems and Applications (IASA),} \\  
{\em P.O. Box 17214, GR-10024 Athens, Greece}    

\end{centering}

\begin{abstract}

In inclusive P-wave charmonium decay, to cancel infrared divergence 
in the colour singlet contribution requires the inclusion of the
colour octet which becomes degenerate with the former in the infrared 
limit. In the corresponding exclusive decay, such an infrared divergence 
does not exist. On this ground, it becomes doubtful whether the colour 
octet is needed in exclusive reactions. A contradiction in the underlying 
picture, however, would result once all strong decay channels are summed. 
We provide an answer to this question in support of the colour octet with 
theoretical arguments as well as an explicit calculation of 
$\c_J \lra p\bar p$.
\end{abstract}

\vspace{2.0cm}

\begin{flushleft}

\end{flushleft}

\vfill
\eject

\section{Introduction}
\label{sec:intro}

Within the standard hard scattering approach (SHSA) 
of Brodsky and Lepage \cite{bl} for exclusive reactions and the later 
modified version (MHSA) of Botts, Li and Sterman \cite{bs,ls}, 
higher Fock states are usually suppressed due to the necessary
hard momentum exchange for ensuring that the constituents of each 
outgoing hadrons are almost collinear. As a consequence, only the knowledge
of valence Fock states is necessary. However, there are exceptions. 
For example, this might not be true when there are suppression mechanisms 
at work due to helicity conservation, flavor symmetry, higher-wave 
wavefunctions etc.. Indeed, it has been shown in inclusive P-wave charmonium 
decay \cite{bod&bra&lep1,bod&bra&lep2} that the next higher Fock 
state, the so-called colour octet, where the constituents 
of the charmonium consist of the usual charm-anticharm pair and an 
additional constituent gluon, is needed not only for an infrared
safe inclusive decay rate but also because they are of the same order
within a non-relativistic QCD velocity expansion formalism. 

Turning to exclusive processes, the same octet mechanism should 
also be at work in $\c_J$ decays. However without a manifestation
of an infrared divergence, the necessity of the colour octet 
was never there. In this paper, we shall try to clarify this 
issue. The paper is organized as follows. In the next section, 
we will describe the situation in inclusive decay and provide the 
main motivation for what are to follow. In Sec. \ref{sec:L_barrier}, 
the role of the orbital angular momentum of the quarkonium and its 
effect on the annihilation are explained in a language consistent 
with the calculation scheme we use below. 
Then power counting arguments for the colour octet is provided in
Sec. \ref{sec:pc}. The remaining sections are dedicated to
explicit calculations. First we will show in Sec. \ref{sec:cs} 
that the colour singlet contribution in $\c_J \lra p \bar p$
within the MHSA is not sufficient, and then the method of calculating
the octet contribution within the SHSA will be explained in 
Sec. \ref{sec:cs+co}. The results of combining the singlet and octet 
contribution within the latter framework will be given and compared
with experiments.

\section{Is Colour Octet Necessary In Exclusive Decay?}
\label{sec:co_needed}

In the early 70's, inclusive decay of charmonium was studied
by Barbieri et al \cite{bgk,bgr} in a series of papers. The family 
of quarkonia, which consists of members made up of bound state of 
heavy quark-antiquark pair, are special because they can be treated 
non-relativistically. These systems decay strongly via annihilation
into gluons. The P-wave charmonia, which are our central interest in
this paper, decay through two gluons at leading order in $\a_s$ 
(\fref{f:incl} (a)). In the rest frame, two gluons 
depart back-to-back with their kinematics completely fixed. The decay 
probability amplitudes are therefore regular, free from any divergence 
as a result. This is the case for $\c_0$ and $\c_2$ at leading order. 
For $\c_1$, the same decay into two massless spin-1 gluons is forbidden. 
The leading process in this case is the decay into a gluon and a 
quark-antiquark pair. A three-body final state is less restrictive and has 
more kinematic freedom. In particular, the quark and antiquark can now depart 
near back-to-back leaving the gluon to take the remaining energy-momentum
and thus ensuring the latter two are conserved. This means that there is no 
lower limit to how soft the gluon can be. It is well known that there is 
an infrared divergence in the tree graph as shown in \fref{f:incl} (b),
that is, provided that all external legs are put on the mass-shell. 
\bfi
\centerline{
\epsfig{figure=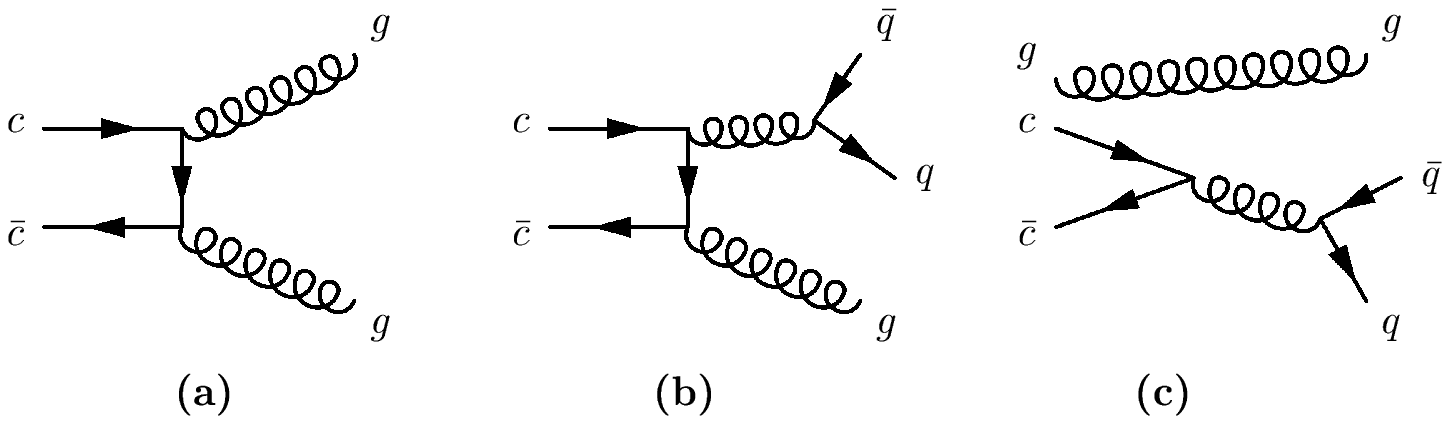,width=5.0in}
}
\caption{The leading decay process for (a) $\c_0$ and $\c_2$
and (b) $\c_1$. A possible decay of the higher Fock state, the 
colour octet, of the $\c_J$ is shown in (c). In the infrared limit 
of the outgoing gluon, (b) and (c) are degenerate.}
\label{f:incl}
\efi
In view of the divergence, it was decided in \cite{bgk,bgr} that
since the $c\bar c$ was really coming from a bound state so it was
perhaps more appropriate for the heavy quarks to be off-shell.
This off-shellness must correspond to the bound state energy $\e$.
With this as a cutoff, the inclusive decay width acquires a logarithmic
$\e$ dependence
\be  \Gamma^{\srm{incl.}}_{\c_1} \sim \a_s^3 \int_\e {{\dd q} \over q}
     \simeq \a_s^3 \log \left (\frac{1}{\e} \right )  \; .
\ee
The same problem actually occurs also in the inclusive decay of
$\c_0$ and $\c_2$ but at next-to-leading order. This form of the
logarithmic dependence on the binding energy of the width was left 
alone until the 90's. The emergence of the colour octet higher Fock
state of charmonium and the realization of its importance to
P-wave charmonium enable finally the cancellation of this 
infrared divergence \cite{bod&bra&lep1,bod&bra&lep2}. 
In the infrared limit of the emitted gluon, the singlet decay graph
becomes degenerate with the annihilation of a $c\bar c$ in a colour
octet state into $q\bar q$ with an accompanying soft spectator gluon.
This is illustrated in \fref{f:incl} (c). The infrared divergence
cancels between \fref{f:incl} (b) and (c). Thus in inclusive decay
because of infrared divergence, it is incorrect to separate the
valence colour singlet state from the higher colour octet state for a
P-wave charmonium. Both have to be included to ensure infrared safe
calculations. The same, however, cannot be said in exclusive decay.
By stipulating what final states should be observed, the infrared
problem is removed. The kinematics of decaying from a P-wave charmonium
bound state into other bound states does not allow any infrared 
divergence. Since no problem was encountered, it was traditional
to calculate the partial decay widths using only the colour singlet
valence Fock state \cite{cz1,andrik,dam&tso&berg,coz2,murg&melis2}
and claims were made that these calculations could give results in
agreement with experimental measurements. In view of the development
in inclusive decay, it is unavoidable to ask the question whether 
colour octet is needed if one concentrates on just a few decay channels. 
If the answer is no to each and every channel, then there would be 
a contradiction since the partial width of each individual channel
could in principle be worked out separately and the sum of them must agree 
with the total hadronic width. In practice, if the requirement of the
colour octet does not reveal itself in at least one of the few-body decay 
channels, it would be hard to find it in the channels with many final 
products. To confirm this contradiction would then be very daunting. 
However it is hard to imagine how the colour octet would 
be important in many-body decays but not in the few-body channels since
the octet comes in already at the leading order of $\c_1$ and at the
next-to-leading order of $\c_0$ and $\c_2$ in inclusive process. With the 
lack of the same infrared divergence, one needs a more compelling reason 
than the less tangible claim of a contradiction to introduce the colour 
octet in exclusive decay. We will provide an argument to this effect 
in the following two sections.

\section{Angular Momentum Barrier To Annihilation}
\label{sec:L_barrier}

Charmonium decays strongly via annihilation into gluons. The heavy mass
of the $c\bar c$ system means that the heavy quark and antiquark tend
to come close to each other when the annihilation is taking place.
The probability of small spatial separation of the $c\bar c$ will
then affects the likelihood of the annihilation to take place.
For a S-wave charmonium, the probability of spatial distribution
is largest at small distance so it is favorable for the annihilation.
But for a P-wave, the $L=1$ orbital angular momentum of
the $c\bar c$ forces them apart and the probability for the heavy
fermions to be near each other tends to be very small and, in fact,
vanishes at zero separation. For this reason, the annihilation of
a P-wave charmonium into gluons is suppressed by this angular momentum
barrier. 

To get an idea of the actual size of the suppression, one only has to
examine the charmonium wavefunction, which contains all the
information of the spatial distribution of the bound state system.
In the case of a S-wave, annihilation occurs at a length scale 
$l\sim 1/M$, the inverse of the charmonium mass. The probability
amplitude for the annihilation then depends on 
$\J_S (l\sim 1/M) \simeq \J_S (l\sim 0)$ the wavefunction at the origin.
As for a P-wave, because of angular momentum $\J_P(0) = 0$, 
so the product of $l$ and the first derivative of the wavefunction has 
to be used instead $\J_P(l\sim 1/M) \simeq l\;\J'_P(l\sim 0)$. 
Fourier transforming these wavefunctions into momentum space, we have
\bea \mbox{S-wave: \hskip 1.0cm} \J_S (0) \mbox{\hskip 0.30cm}
     & \longrightarrow & \tilde  \J_S (k)                  \nonumber \\
     \mbox{P-wave: \hskip 1.0cm} l\;\J'_P(0)
     & \longrightarrow & \frac{k}{M} \tilde \J_P (k)  \; , \nonumber
\eea
where $k$ is the internal transverse momenta of the quarkonium of the 
order of a few hundreds MeV. It is now obvious that the suppression in 
the P-wave wavefunction is of the form $1/M$ which is substantial for 
a heavy charmonium to decay through annihilation into gluons.

\section{Power Counting in the Charmonium Mass in The Decay Probability Amplitude}
\label{sec:pc}


In the last section, we showed that P-wave charmonium annihilation
was unfavorable in comparison to S-wave because of the angular
momentum barrier. To answer the question we posed earlier, it remains to 
show how this fact brings about the need of the colour octet. 
To do that we need to examine the large scale dependence of the 
decay probability amplitude or more specifically to perform power counting
using that scale. The simplest scheme to work out a decay amplitude is
via the hard scattering scheme already mentioned at the beginning. 
In this SHSA method, the decay amplitudes for a charmonium 
$C =J/\J, \c_J, \dots$ to decay into a pair of pseudo-scalars 
$\cp = \p, K, \dots$ etc. and a proton-antiproton pair $p \bar p$ have 
the form
\be \cm_{C \lra \cp \cp} \sim f_{C} \f_{C}  \ot 
    f_\cp \f_\cp   \ot f_\cp \f_\cp   \ot T_H   
\label{eq:ha_pseu}
\ee
and
\be \cm_{C \lra p \bar p} \sim f_{C} \f_{C}  \ot 
    f_p \f_p  \ot f_{\bar p} \f_{\bar p}  \ot T_H   \; ,
\label{eq:ha_prot}
\ee
respectively. In \ederef{eq:ha_pseu}{eq:ha_prot}, the amplitudes
are each given by a convolution over probability amplitudes $\f$'s and 
the hard perturbative part $T_H$. The partial decay width in each case is 
given by $ \Gamma \sim |\cm |^2/M $. So $\cm$ is of mass dimension one.
In these amplitudes, only the decay constants $f$'s and $T_H$ carry a mass 
dimension of some power. These must altogether make up for the right
mass dimension of $\cm$. Each product $f \f$ is nothing but
the internal transverse momenta integrated light-cone wavefunction
\be  f \f (x) = \int^Q \prod^{N-1}_i \intksno{i} \; 
     \J (x;\Kpn{1},\dots,\Kpn{N-1})
\label{eq:prob_amp}
\ee
where $Q$ is some large cutoff scale, given approximately by the
large scale of the hard process in question, for the $N-1$ internal momentum
integrals over the $N$ particle wavefunction $\J$, and $x$ represents
collectively ($x_1,\dots,x_N$) the light-cone fractions. The normalization
of the wavefunction is given by
\be  \int \prod^{N-1}_i \dd x_i \intksno{i} \; 
     \left | \J (x;\Kpn{1},\dots,\Kpn{N-1}) \right |^2 = P \; .
\label{eq:wfun_norm}
\ee
Here $P$ is the probability of this particular hadron to be in
this state with $N$ constituents and is of course a dimensionless number. 
Using \ederef{eq:prob_amp}{eq:wfun_norm}, the mass dimension of the decay
constant of any hadrons can be deduced. In this paper, we restrict our 
considerations to only charmonium states with up to three constituents for 
any charmonium. In Table \ref{tab:mass_dim}, the mass dimensions of the decay 
constants together with the number of constituents of several hadrons are 
listed. The odd ones in the table are the decay constants of the colour singlet 
$\c^{(1)}_J$, which has a dimension of two although it is coming from 
a 2-particle wavefunction, and the colour octet ${J/\J}^{(8)}$, which
is a 3-particle wavefunction but has a dimension of three.
The reason for this is that as we have already discussed, 
the P-wave wavefunction vanishes at the origin so the product of 
the first derivative of the wavefunction at the origin and 
the annihilation size $l$ must be used instead. This has the effect 
of increasing the mass dimension of the decay constant by one. 
In the case of ${J/\J}^{(8)}$, the fact that $J/\J$ is a $1^{--}$ 
charmonium excludes the possibility of the leading colour octet contribution 
to be in a S-wave three-particle wavefunction, which must be found in a 
three-particle P-wave state instead. Note that for the three-constituent
colour octet state of the $J/\J$, the $c\bar c$ pair itself can be in $^3P_J$ 
or $^1S_0$ as is frequently considered in inclusive $J/\J$ productions
\cite{ben&rot,bra&che,fle&mak,tan&van},
but the full wavefunction must have an orbital angular momentum
of one. In Table \ref{tab:qno}, the orbital angular momentum $L^c_{\bar c}$
and spin $S_{c\bar c}$ of the $c\bar c$ pair, the orbital angular 
momentum between the pair and the constituent gluon $L^{c\bar c}_g$, 
the number of constituent gluon $N_g$, the parity $\cp$ and C-parity $\cc$ 
of each state of the $J/\J$ and $\c_J$ up to three constituents are 
tabulated. This shows that the $^1S_0$ colour octet state of the $J/\J$,
although has no orbital angular momentum in the $c\bar c$ pair, must have
one unit of this momentum between the pair and the constituent gluon
otherwise this state cannot be in the required $1^{--}$ state of the $J/\J$.
So although the heavy quark pair is in an S-wave, the three-particle
wavefunction is effectively in a P-wave. That is why the decay constants 
of the two three-constituent colour octet states of the $J/\J$, usually labelled 
as $^3P_J$ and $^1S_0$, are both of mass dimension three. Therefore the
mass dimension indicated in Table \ref{tab:mass_dim} for $J/\J^{(8)}$ 
is true for both states and $L = L^c_{\bar c}+L^{c\bar c}_g$ in these 
two cases. In Table \ref{tab:qno}, two possibilities of the potentially 
possible three-constituent octet states of the $J/\J$ with the heavy quark
pair in $^3S_1$, but which in reality do not exist hence the attached 
``?'' marks, are shown. The purpose of these two entries in the table
is to show how using the remaining available degree of freedom 
$L^{c\bar c}_g =0$ or $1$, it is still not possible to change the states into
a $1^{--}$ because $L^{c\bar c}_g$ only affects $\cp$ and not $\cc$. 
In the case of the octet $^1S_0$, it is the $L^{c\bar c}_g = 1$ which permits
this possibility to exist. 

Since the decay process of $C$ has only one large scale, that is the
charmonium mass $M$, $T_H$ must contain sufficient power of
this scale to make up for the mass dimension of $\cm$. So in 
the following decay processes, the power counting in or the power 
dependence on $\mjp \sim \mcj \sim M$ of the charmonium in the colour 
singlet $\cm^{(1)}$ and octet $\cm^{(8)}$ amplitude with up to three 
constituents in the respective charmonia goes as follows 
\footnote{
Note that our power counting is done by taking the charmonium mass 
$M$ to be the largest and most dominant scale in the process and
the ratios of any quantities with any others are much less important 
than those of $M$ with these quantities. In this sense, one can
forget about the actual magnitude of the other quantities with a mass 
dimension, such as the decay constants, with the understanding that they 
are small in front of $M$. This permits us to concentrate only on the power 
of $M$ in the amplitudes above.}
\footnote{
We have, for the sake of simplicity, ignored the dependence of the 
charmonium decay constants on the large scale $M$ because this is at most
logarithmic and is therefore much weaker than the power dependence.}
\bea 
     \cm^{(1)}_{J/\J \lra \cp \cp} & \sim & 
     \mjp \frac{f^{(1)}_{J/\J}}{\mjp} \Big ( \frac{f_\cp}{\mjp} \Big )^2  
     \sim \frac{1}{M^2}   
\label{eq:js_Ps}              \\
     \cm^{(8)}_{J/\J \lra \cp \cp} & \sim & 
     \mjp \frac{f^{(8)}_{J/\J}}{\mjp^3} \Big ( \frac{f_\cp}{\mjp} \Big )^2  
     \sim \frac{1}{M^4}   
\label{eq:jo_Ps}              \\
     \cm^{(1)}_{\c_J \lra \cp \cp} & \sim & 
     \mcj \frac{f^{(1)}_{\c_J}}{\mcj^2} \Big ( \frac{f_\cp}{\mcj} \Big )^2  
     \sim \frac{1}{M^3}  
\label{eq:cs_Ps}             \\
     \cm^{(8)}_{\c_J \lra \cp \cp} & \sim & 
     \mcj \frac{f^{(8)}_{\c_J}}{\mcj^2} \Big ( \frac{f_\cp}{\mcj} \Big )^2  
     \sim \frac{1}{M^3}  
\label{eq:co_Ps}             \\
               & & \nonumber \\
     \cm^{(1)}_{J/\J \lra p\bar p} & \sim & 
     \mjp \frac{f^{(1)}_{J/\J}}{\mjp} \Big ( \frac{f_p}{\mjp^2} \Big )^2  
     \sim \frac{1}{M^4}  
\label{eq:js_p}               \\
     \cm^{(8)}_{J/\J \lra p\bar p} & \sim & 
     \mjp \frac{f^{(8)}_{J/\J}}{\mjp^3} \Big ( \frac{f_p}{\mjp^2} \Big )^2  
     \sim \frac{1}{M^6}  
\label{eq:jo_p}               \\
     \cm^{(1)}_{\c_J \lra p\bar p} & \sim & 
     \mcj \frac{f^{(1)}_{\c_J}}{\mcj^2} \Big ( \frac{f_p}{\mcj^2} \Big )^2  
     \sim \frac{1}{M^5}   
\label{eq:cs_p}              \\
     \cm^{(8)}_{\c_J \lra p\bar p} & \sim & 
     \mcj \frac{f^{(8)}_{\c_J}}{\mcj^2} \Big ( \frac{f_p}{\mcj^2} \Big )^2  
     \sim \frac{1}{M^5}   \; .
\label{eq:co_p}
\eea
\equref{eq:cs_Ps}{eq:co_Ps}{eq:cs_p}{eq:co_p} show that the probability
amplitude of the colour singlet and octet of $\c_J$ decay depend on the same 
power of $M$. The contribution from the colour octet higher Fock state of the 
P-wave charmonium is not suppressed at all as naively expected in comparison 
with the singlet contribution. However, this proves to be true in the 
case of the S-wave $J/\J$. \equref{eq:js_Ps}{eq:jo_Ps}{eq:js_p}{eq:jo_p}
show that the colour octet contribution to the amplitude of the $J/\J$ 
decay is indeed suppressed by $1/M^2$ (this suppression would have 
been only $1/M$ if the leading three-particle wavefunctions of the 
colour octet states of $J/\J$ were not forced to be in a P-wave when all 
three constituents have been taken into account, as such there is an additional 
factor of $1/M$). The power of $M$ counting of the octet $J/\J$ decay 
amplitudes in \ederef{eq:jo_Ps}{eq:jo_p} are applicable for the $c\bar c$ in 
both the $^3P_J$ and $^1S_0$ state. Comparing the colour singlet decays of the 
$J/\J$ in \ederef{eq:js_Ps}{eq:js_p} with the $\c_J$ decays in 
\ederef{eq:cs_Ps}{eq:cs_p} shows that the $J/\J$ amplitude is larger by a 
power of $M$ in each case. The reason for this is due precisely to the fact 
that the P-wave charmonium wavefunction is suppressed by $1/M$ because of 
the $L=1$ angular momentum as explained in the previous section. Without this 
angular momentum suppression, the colour octet contribution in $\c_J$ decays 
could be neglected. But as it is, provided that there is no other form of 
significant suppression, which is indeed true \cite{bks,bks2,wong}, colour 
octet contribution must be included.

\section{Colour Singlet Contribution in the Modified Hard Scattering
Approach}
\label{sec:cs}

The previous sections provided the arguments for the inclusion of
the colour octet in $\c_J$ decay. The decay into pseudo-scalars
with the octet contribution has been considered in \cite{bks,bks2}. In this 
section, we will show by explicit calculation, that the size of the colour 
singlet contribution to the decay into proton-antiproton is small in comparison 
with the experimental partial width. The scheme that we will use is the 
MHSA because of two advantages it has over the SHSA. The first is the 
renormalization scales of the $\a_s$'s are determined dynamically by the 
virtualities that flow in the diagrams that represent the decay process. 
The second is the problematic endpoint regions of the distribution 
amplitudes where the virtualities tend to approach $\lqcd$ are suppressed
by the inclusion of the Sudakov factor $\exp \{-S\}$. The decay probability 
amplitude takes the form
\be  \cm_{\c_J \lra p\bar p} \sim \J_{\c_J} \ot \J_p 
     \ot \J_{\bar p} \ot T_H(\ot \a_s) \ot  \exp \{-S \}   
     \; .
\label{eq:mhsa}
\ee
The $\a_s$ in $T_H$ is expressed in such a way to signify that the coupling,
although hidden within the hard part, is part of the convolution and is not 
taken outside as a prefactor as in the SHSA. As can be seen, a proton
wavefunction has to be used. This is potentially problematic because
of all the uncertainties surrounding the proton distribution amplitude. 
Although proton model wavefunctions exist via QCD sum rule derivation
(see any of the refs. \cite{cz1,cz,coz,gs,ks,bst}), none of them
could describe the magnetic form factor below $Q^2 \sim 50$ GeV$^2$
in a self-consistent manner \cite{rad,bjkbs}. Fortunately by relinquishing
a formal derivation, one can turn to a phenomenological approach. 
Since the high momentum tail of the wavefunction is clearly not dominant
in present day experimental accessible regions --- $Q^2$ even up to $70$--$100$ 
GeV$^2$ are still nowhere near the asymptotic region as far as the proton 
is concerned --- the soft part of the wavefunction must still be dominant 
up to this range. Then by fitting the soft overlap of the initial and final 
proton wavefunction, or the so-called soft Feynman contribution to the 
magnetic form factor, to experimental constraints such as the 
$J/\J \lra p \bar p$ decay width, proton valence 
quark distribution and the magnetic form factor measurements, a model 
proton wavefunction can be constructed \cite{bolz&kroll1}. This, by construction, 
works automatically in the $Q^2 \sim $ 10 GeV$^2$ region, which is precisely 
the region of interest in this paper. We stress that the problem discussed
here concerning the proton wavefunction is strictly limited to this
sector of the convolution in \ederef{eq:ha_pseu}{eq:ha_prot}.
It does not affect the charmonium or the perturbative hard part because
as in the problem of the proton electromagnetic form factor, it involves only 
the proton wavefunction. Once we used the phenomenologically constructed 
wavefunction and not those from, for example, sum rules in this sector, the 
problem is taken care of. 

The proton wavefunction is given by
\bea
%
  \label{pstate}
  |\,p,+ \,\ran\, &
=
  \frac{\varepsilon _{a_{1}a_{2}a_{3}}}{\sqrt{3!}} 
  \int
  [{\rm d}x]
  [{\rm d}^{2}{\Kp}] &
  \Bigl\{ 
         \j^p_{123}\:|\,u_+^{a_1} u_-^{a_2} d_+^{a_3}\,\ran
       + \j^p_{213}\:|\,u_-^{a_1} u_+^{a_2} d_+^{a_3}\,\ran
  \phantom{ \Bigr\} \;\;\; }
\nonumber \\
   & & \mbox{\hskip 2.0cm} 
       - \Bigl(\j^p _{132}\, + \,
         \j^p_{231}\Bigr)\:|\,u_+^{a_1} u_+^{a_2} d_-^{a_3}\,\ran
  \Bigr\} \;\;\; 
\eea
where the decay constant $f_p$ and the distribution amplitude 
$\f^p_{123}$ in the scalar function
\be 
  \j_{123}(x,\Kp) =  \j (x_1,x_2,x_3;\Kpn 1,\Kpn 2,\Kpn 3) =
  \frac{1}{8\sqrt{3!}} \,  f_p (\mu_F)
  \phi^p_{123} (x,\mu_{F})\,  \Omega_p (x,\Kp) \:. 
\label{Psiansatz}
\ee
depend on the factorization scale $\m_F$, and the two measures in square
brackets are the usual ones
\be
  \dxb = \prod_{i=1}^3 {\rm d}x_i\,\delta(1-\sum^3_{i=1} x_i) \qquad
  \dkb = \frac{1}{(16\pi^3)^{2}}\,\prod_{i=1}^3 
     {\rm d}^2 \Kpn{i} \, \delta^{(2)}(\sum^3_{i=1} \Kpn{i}) 
     \:.
\ee
The internal transverse momentum dependent part can best be expressed
as a Gaussian
\be 
  \Omega_p(x,\Kp) =
  (16\pi ^{2})^{2}  \frac{a_p^{4}}{x_{1}x_{2}x_{3}}
  \exp
     \left [
            -a_p^{2} \sum_{i=1}^{3} {\Kpn{i}}^{2}/x_{i}
     \right ]\:.
\label{BLHMOmega}
\ee
where $a_p = 0.75$ GeV$^{-1}$ is the transverse size parameter and 
$f_p (\m_0) = 6.64 \times 10^{-3}$ GeV$^2$ at the reference scale 
$\m_0 = 1.0$ GeV. These values are obtained in the construction in
\cite{bolz&kroll1}.
The proton distribution amplitude in terms of the Appell polynomials
$\tilde \f^n_{123}$'s, the scale dependent expansion coefficients 
$B^p_n(\m_F)$ and the asymptotic distribution amplitude $\f_{\rm AS}(x)$ is
\bea
  \f^p_{123}(x,\m_F) &=& \f_{\rm AS}(x) \left[1 +
  \sum_{n=1}^{\infty} B^p_n(\mu_F)\,\tilde \f^n_{123}(x,\mu_F) \right]
  \\
  \f^p_{123}(x,\m_0) &=& \f_{\rm{AS}}(x) 
      \left [1 + \frac{3}{4} \tilde \f^1_{123}(x)
               + \frac{1}{4} \tilde \f^2_{123}(x) \right ]
      \;\;\; = \; 60 x_1 x_2 x_3 [1+3 x_1]  \; .
\label{DAentw}
\eea
\eref{DAentw} shows the rather simple form of $\f^p_{123}$ at the scale
$\m_0$. For the scale dependent expression of $B^p_n(\m_F)$ and 
$f_p(\m_F)$ and those of the more general baryon wavefunctions, we refer 
the readers to our other paper \cite{wong}.

\bfi
\centerline{
\epsfig{figure=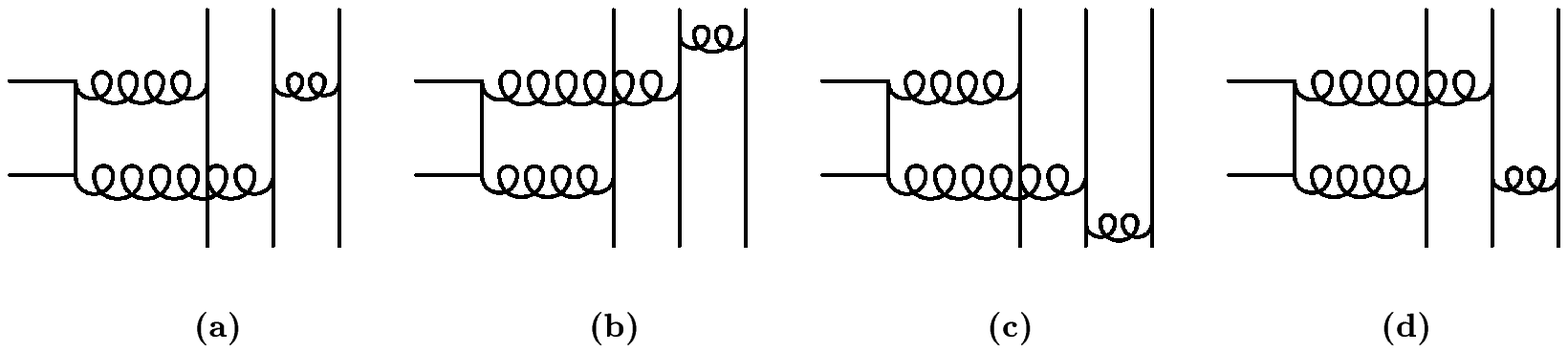,width=5.0in}
}
\caption{Feynman graphs for the colour singlet $\c_J$ decay into
proton-antiproton.}
\label{f:sing}
\efi

Equipped with this proton wavefunction, we are ready to work out the 
perturbative hard part $T_H$ in MHSA. Because of C-parity conservation 
\cite{nov&etal}, the colour singlet $\c_J$ component can only annihilate 
into proton-antiproton through two gluons \cite{wong}. The Feynman graphs 
are depicted in \fref{f:sing}. The three light quark lines in each case can 
be permuted further to give more graphs but these can be included by a 
symmetry factor. Basing on \fref{f:sing} (a), $\hat T^J_H(x,y,\B,\B')$ in 
transverse separation space $\B$ and $\B'$ of the proton-antiproton and in 
terms of 
\bea 
    \cc^J_1 (x,y) &=& y_3 \; (1-2 x_1)^{2-J}    \\
    \cc^J_2 (x,y) &=& (-1)^{J-1} \; y_3 \; x_1 (x_1-y_1) 
\eea
worked out to be \cite{wong}
\bea  \hat T^J_H (x,y,\B,\B') &=& 
      {{2^{9} \sqrt{2} \; m_c} \over {9\sqrt{3}} } \; \;
      \a_s(t_1) \a_s(t_2) \a_s(t_3) 
      \; \; \d^2 (\B_{1}-\B_{1}'-\B_{3}+\B_{3}') 
      \nonumber \\ \fx & & \fx \times
      \frac{i\p}{2} H^{(1)}_0 (\sqrt{x_3 y_3} (2 m_c) |\B_3|) \; \;      
      \frac{i\p}{2} H^{(1)}_0 (\sqrt{(1-x_1) y_3} (2 m_c) |\B_1-\B_1'|)
      \nonumber \\ \fx & & \fx \times
      \Bigg \{ 
      i\p \bigg [ {{H^{(1)}_0 (\sqrt{x_1 y_1} (2 m_c) |\B_1'|)}
                  \over {(x_1+y+1)(1-x_1-y_1)}}
               \Big ( \cc^J_1(x,y) 
                     + {{\cc^J_2(x,y)} \over {x_1+y_1}}
               \Big )       \nonumber \\ \fx & & \; \; \; \; \;
                -{{H^{(1)}_0 (\sqrt{(1-x_1)(1-y_1)} (2 m_c) |\B_1'|)}
                  \over {(1-x_1-y_1)(2-x_1-y_1)}}
               \Big ( \cc^J_1(x,y) 
                     + {{\cc^J_2(x,y)} \over {2-x_1-y_1}}
               \Big )       
          \bigg ]           \nonumber \\ \fx & & 
              -\frac{4}{\p} 
                 {{K^{(1)}_0 (\sqrt{x_1 (1-y_1) + y_1 (1-x_1)} m_c |\B_1'|)}
                  \over {(x_1+y_1)(2-x_1-y_1)}}  \nonumber \\ 
              \fx & & \;\;\;\; 
               \Big ( \cc^J_1 (x,y) 
                     + {{2 \; \cc^J_2 (x,y)} 
                        \over {(x_1+y_1)(2-x_1-y_1)}}
               \Big )       \nonumber \\ \fx & & \;
      -{{2 m_c |\B_1'| \; \cc^J_2(x,y) \; 
         K^{(1)}_1(\sqrt{x_1 (1-y_1) + y_1 (1-x_1)} m_c |\B_1'|)}  
        \over {\sqrt{x_1 (1-y_1) + y_1 (1-x_1)} \; (2-x_1-y_1) (x_1+y_1)}}
      \Bigg \}    \nonumber \\ \fx & & 
       + \Big ( (x, \; \B) \llra (y, \; \B') \Big ) \; \; \; .
\label{eq:T_H}
\eea
The $H^{(1)}_0$ is the Hankel function, $K^{(1)}_0$ and $K^{(1)}_1$ 
are the modified Bessel functions of the second kind. In the above
and throughout this paper, we use the one-loop expression for 
$\a_s$ and set $\lqcd = 220$ MeV and $m_c = 1.5$ GeV. 
The renormalization scales $t_1, t_2$ and $t_3$ in the product 
of three $\a_s$'s here are determined by the largest of either 
the virtualities in the neighborhood of the vertices associated with each 
$\a_s$ or the smallest transverse separation $\B$ or $\B'$ of the 
constituents of the proton or antiproton in accordance to the method 
described in \cite{bs,ls,bjkbs}. 

Convoluting this with the wavefunctions and the Sudakov factor 
gives the decay form factor
\bea \cb_J^{p \; (1)} 
    &=& -i {{\sqrt{3} |R_p'(0)| \; \s_J} \over {4\sqrt{\p} m_c^{3/2}} }
    \int \dxb \dyb \; \intbs{1} \intbs{3} \intbds{1} \intbds{3}
    \; \hat T^J_H (x,y,\B,\B')    \nonumber \\
    & \times & 
  \bigg \{\; 
          \hat\j^{p}_{123}(x,\B) \hat\j^{p}_{123}(y,\B')  
         +\hat\j^{p}_{321}(x,\B) \hat\j^{p}_{321}(y,\B')  
          \nonumber \\
  \fx & &  +
      \Big(\hat\j^{p}_{123}(x,\B ) + \hat\j^{p}_{321}(x,\B ) \Big)\! 
      \Big(\hat\j^{p}_{123}(y,\B') + \hat\j^{p}_{321}(y,\B') \Big)    \nonumber \\ 
  \fx & &  +
   (2 \llra 3 )
  \bigg \} \;\; \exp [-S(x,y,\B,\B',2m_c)] 
\label{eq:s_df}
\eea
where $\s_J = 1/\sqrt{2}, 1$ for $J=1,2$ respectively, and the first
derivative of the P-wave radial wavefunction of $\c_J$ is well known
$|R_p'(0)| = 0.22$ GeV$^{5/2}$. The form of the Sudakov function $S$
can be found in, for example, ref. \cite{bks2}. The Sudakov exponential 
factor performs a very important function of suppressing the regions
where any of the virtualities $t_i$ approaching the dangerous scale of
$\lqcd^2$. It renders the calculation self-consistent in the sense that
the size of the contribution from the soft region can never be the
bulk of the hard part. With this self-consistency in place,
the decay form factor is related to the decay width by
\be
   \Gamma (\c_1\rightarrow p\bar p) 
   = { {\r_{\rm p.s.}(M_p/\mcjo)} \over {16 \p \mcjo} } \; 
     \frac{1}{3} \; \sum_{\l's} \left |\cm^1_{\l_1 \l_2 \l} \right |^2
   = { {\r_{\rm p.s.}(M_p/\mcjo) \, m_c^2} \over {3 \p \mcjo} } \; 
     \; \left | \cb^p_1 \right |^2  \; ,
\label{eq:gam_1}
\ee
and
\be
   \Gamma (\c_2\rightarrow p \bar p) 
   = { {\r_{\rm p.s.}(M_p/\mcjt)} \over {16 \p \mcjt} } \;
     \frac{1}{5} \; \sum_{\l's} \left |\cm^2_{\l_1 \l_2 \l} \right |^2 
   = { {\r_{\rm p.s.}(M_p/\mcjt) \, m_c^2} \over {10 \p \mcjt} } \; 
     \; \left | \cb^p_2 \right |^2  \; .
\label{eq:gam_2}
\ee
Since in the calculation proton is taken to be massless,
phase space has been corrected by $\r_{\rm p.s.}(z)= \sqrt{1-4 z^2}$.


\bfi
\centerline{
\epsfig{figure=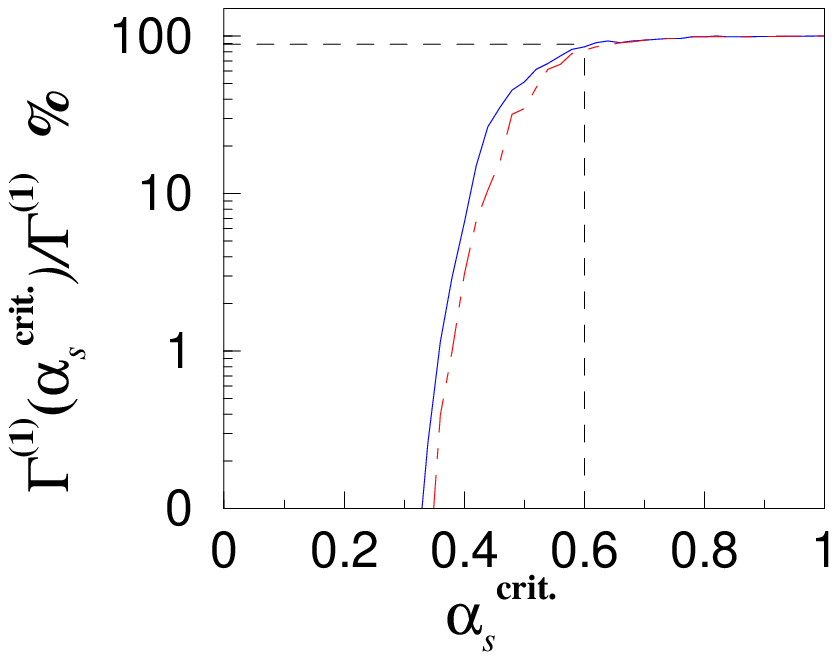,width=2.00in}
}
\caption{Percentage of the colour singlet contribution to
the $\c_J \lra p\bar p$ decay width that comes from
region of values of $\a_s$ below some critical cutoff value 
$\a_s^{\srm{crit.}}$. The solid and dot-dashed line are for $\c_1$
and $\c_2$, respectively. This shows that the bulk of the
contributions in fact comes from $\a_s < 0.6$ region.} 
\label{f:alpha_crit}
\efi

Our results are shown in Table \ref{tab:sing_nucl} together with
the experimental measurements from Particle Data Group (PDG) \cite{pdg} and 
the more recent BES Collaboration \cite{bes}. The errors associated
with the PDG results are shown but those associated with BES 
are not because they are too large. One can consult them in
ref. \cite{bes}. The results reveal a large discrepancy between the colour 
singlet contributions and the experimental widths.  
Before drawing any definite conclusions from these results,
we first show that our calculation within MHSA is perturbatively consistent.
To do that, an artificial cutoff $\a_s^{\srm{crit.}}$ in the value of the 
coupling was introduced so that the integrations in \eref{eq:s_df} include
only regions where $\prod^3_{i=1} \a_s(t_i) < (\a_s^{\srm{crit.}})^3$
was true. The value of $\a_s^{\srm{crit.}}$ was varied to determine
how much of the final results came from hard regions.
The percentage of the final widths are plotted against 
$\a_s^{\srm{crit.}}$ in \fref{f:alpha_crit} for both $J=1$ and $J=2$
charmonium. It shows clearly that by far the main contributions come from
hard regions in which $\a_s$ remains below 0.6. So the Sudakov factor
performed as expected to keep the soft non-perturbative region under
control and suppressed.  

To assess the uncertainties in the results, we have used the freedom from 
the uncertainty of the relation between $R_p'(0)$ and $m_c$ from fit to 
charmonium parameters \cite{mag&pet} and from quarkonium potential model 
\cite{qui&ros} and also the value of $\lqcd$ to maximize our results to see 
if they can be raised up to the experimental results. We found that they
maximized with values of $R_p'(0) =$ 0.194 GeV$^{5/2}$ and $m_c =$ 1.35 GeV
and a larger $\lqcd =$ 0.25 GeV. The colour singlet contribution to the
decay width $\Gamma^{(1)}$ remains for $J=1$, 5--9 times and for $J=2$,
2--4 times below the measured results. 
At this point, one may wonder that the uncertainties of the proton 
distribution amplitude or wavefunction may make up for the remaining 
difference. As we have already discussed above, many proton wavefunctions 
derived from QCD sum rules do not work at our present scale of interest. 
For this very reason, the particular one we used here was constructed
phenomenologically. Thus the apparent many choices of wavefunction and 
the fact that using each and everyone of them may lead to very different 
results are illusionary. One simply cannot use these other wavefunctions 
for $\c_J$ decays. As to the nucleon decay constant, other values do exist
from QCD sum rules or from lattice. For example, the values from
QCD sum rules is about $5.0 \times 10^{-3}$  GeV${^2}$ 
\cite{murg&melis2,coz} or from lattice QCD there exist the two values 
$2.9$ or $6.6 \times 10^{-3}$ GeV${^2}$ \cite{bolz&kroll1}. Note that 
these are all smaller than ours, so using them will give even smaller 
results for the singlet contribution and not larger. The large 
experimental discrepancy of the data between PDG and BES cannot be 
explained but assuming that the true widths are somewhere in between or even 
near the BES results, it is clear that the colour singlet contribution alone 
cannot account for the experimental partial decay widths in agreement with our 
theoretical arguments presented in Sec. \ref{sec:L_barrier} and \ref{sec:pc}.

\section{Combining Colour Singlet and Octet Contribution in the Standard
Hard Scattering Approach}
\label{sec:cs+co}

In the last section, we have shown that colour singlet contribution
is not enough to account for the experimental measurements contradicting
all earlier similar calculations done before the emergence of the colour octet 
state in P-wave charmonium and the associated improved understanding of the 
role of the higher Fock state of the quarkonium. It remains to show that
the colour singlet and octet combination do indeed provide reasonable 
agreement to experimental data. We will outline the method of 
dealing with the colour octet and describe our calculations. The combined
results will be shown but the details are given elsewhere \cite{wong}.

The colour singlet calculation above was performed within the MHSA because
of the two already mentioned advantages. The drawback of this scheme is,
however, that it is more complicated to do a calculation. One has to 
include full wavefunctions and not only the distribution amplitudes. This
means internal transverse momenta have to be kept in the hard perturbative 
part $T_H$. For the colour singlet contribution to $\c_J$ decay, this proves 
to be manageable. We have seen the transverse separation space version
of $T_H$ in \eref{eq:T_H}. It contains several hyper-geometric functions
and is therefore non-trivial. In the case of the colour octet, we will see 
presently that the calculation involves many more diagrams. In addition, proton 
and antiproton each carries three valence quarks. All these add up to the
dimensions of the numerical integrations and complicate the calculation
much more. The previous calculation was done within the MHSA to avoid
the ambiguity of having to choose a renormalization scale by hand for $\a_s$ 
since the decay widths depend on the sixth power of the coupling $\a_s^6$
which leaves too much freedom or there is a lack of constraint to arrive
at results that agree with data. In view of the complexity of MHSA, we 
will keep the calculation simple and avoid tangling ourselves in MHSA 
in the high dimensional computations by retreating back to the SHSA. 
This will, of course, bring back all the shortcomings of the SHSA
but in spite of these, the SHSA can often be used to make a reasonable
first estimate of the results.

As is now well known, the colour octet state of a P-wave charmonium 
includes, in addition to the usual $c\bar c$, a constituent gluon. The
presence of this in the initial state requires it to be properly 
``put away'' in the final hadrons in order not to have a net colour.
This colour neutralization can be achieved in two different ways. The
first is to allow this gluon to enter directly into or be a constituent
in the final wavefunction of one of the outgoing hadrons. An example 
is shown in \fref{f:col_neu} (a). SHSA requires that no participant in the 
hard interaction to be disconnected from the rest as a consequence
of the collinear approximation in SHSA. That is all constituents of
any hadrons, be they part of an incoming or outgoing hadron, must be 
almost collinear. Therefore the connectedness of the constituents
will ensure the even distribution of the hard momenta amongst
the final constituents. The alternative is to let the constituent gluon 
be part of the hard perturbative part, that is to let it enter and end 
in $T_H$. \fref{f:col_neu} (b) shows this alternative. Comparing
\fref{f:col_neu} (a) and (b), it is clear that the first is down by
$1/\mcj$ assuming hard momenta run through all connecting lines in $T_H$ as 
is usual in SHSA. Moreover, by including a higher Fock state in one of
the outgoing hadrons will lead to extra suppression due to the smaller
probability of that state, not to mention the undesirable involvement
of further unknown wavefunction. The second possibility will be our choice
for colour neutralization in the rest of this paper. 

\bfi
\centerline{
\epsfig{figure=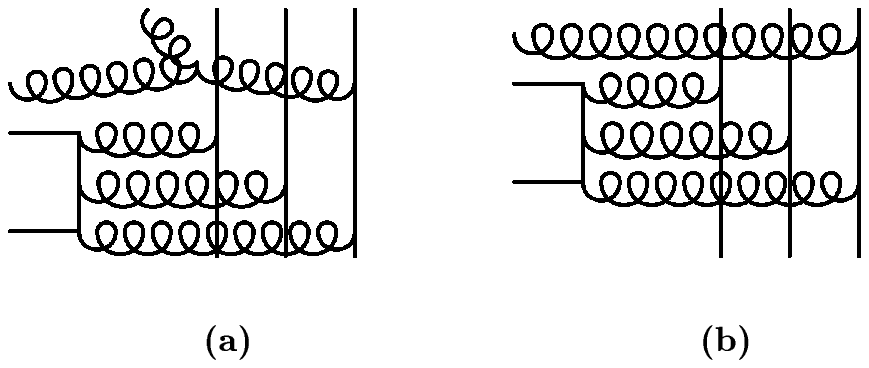,width=3.50in}
}
\caption{Example of two ways to achieve colour neutralization:
(a) the constituent gluon enters the nucleon as part of the constituent
in the nucleon after exchanging a gluon with the other constituents, 
(b) this gluon ends in the perturbative hard part $T_H$.}
\label{f:col_neu}
\efi

\bfi
\centerline{
\epsfig{figure=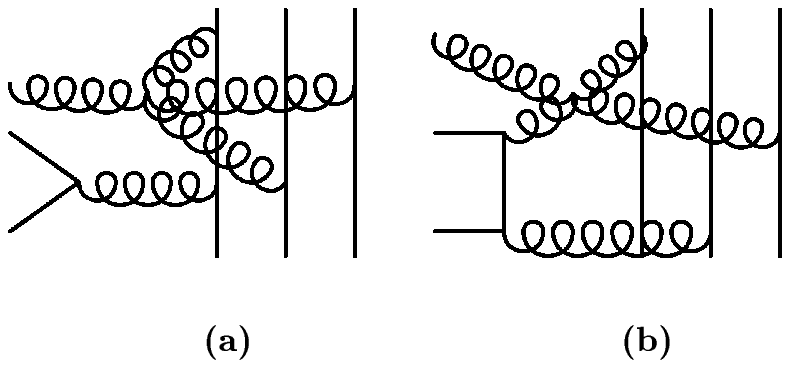,width=3.0in}
}
\caption{Example of $c\bar c$ annihilates through (a) one and (b) two gluons.
Graph (a) is impossible in the colour singlet contribution.}
\label{f:ot_g}
\efi

As we mentioned earlier, C-parity only permits the $c\bar c$ to annihilate
through two gluons in the colour singlet contribution. On the contrary in
the colour octet, the very presence of a constituent gluon in the initial 
state enables many more possibilities. The heavy quark-antiquark pair, which 
is now protected from the C-parity constraint by the constituent gluon,
can annihilate through one, two and three gluons. \fref{f:col_neu} (b)
is an example of the latter and \fref{f:ot_g} (a) and (b) show a
one-gluon and a two-gluon possibility, respectively. \fref{f:ot_g} (a)
and \fref{f:col_neu} (b) are not permitted in the singlet case.
In all, the graphs can be divided into ten groups based on ten basic
graphs from which the members of each group can be constructed. 
These basic graphs are shown in \fref{f:fig_basic}. The four graphs 
in \fref{f:sing} are in fact counted as one basic graph of group (3) since 
they are related by symmetry. The members of all groups are generated by 
using the afore-mentioned colour neutralization method. That is to let the 
constituent gluon end in all possible allowed position on the basic graph 
as part of $T_H$. This generates between four to eleven graphs 
depending on the group. For example, \fref{f:col_neu} (b) is generated
from group (5) by adding a gluon emerging from the $\c_J$ and ending
on one of the light quark lines, \fref{f:ot_g} (a) is from group (10) 
and \fref{f:ot_g} (b) is from group (4) of \fref{f:fig_basic}. 
The latter two are generated by letting the constituent gluon to end on the
three-gluon vertex on the group (10) and (4) basic graph, respectively. 
More details on these are given elsewhere \cite{wong}.  

\bfi
\centerline{
\epsfig{figure=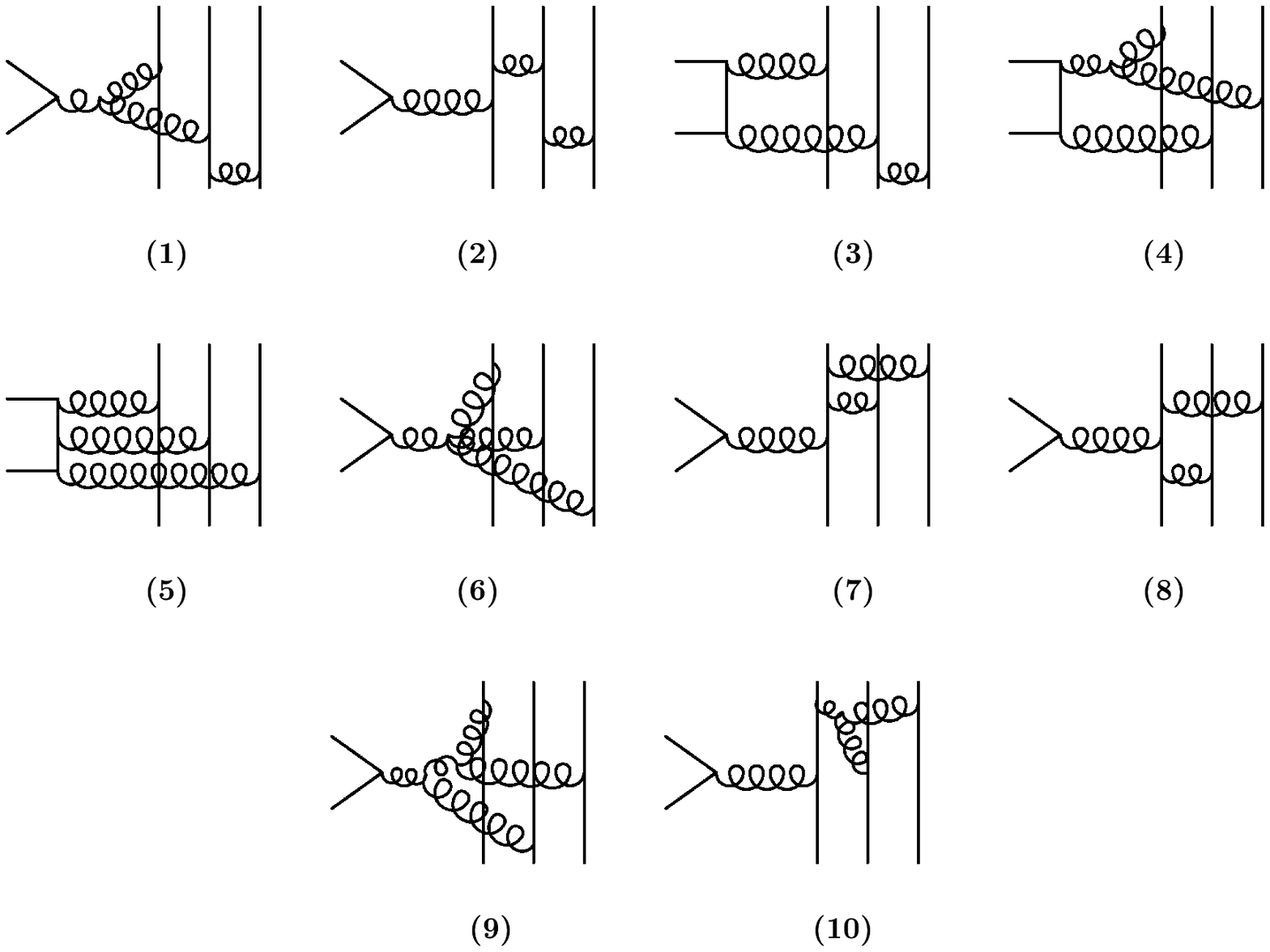,width=5.50in}
}
\caption{The feynman graphs that represent the ten basic groups from
which the graphs of the colour octet contributions are generated. Each graph
is assigned a group number as indicated.}
\label{f:fig_basic}
\efi


The decay amplitude has the form of \eref{eq:ha_prot}. The convolution
now involves only distribution amplitudes. That of the proton
is shown in Sec. \ref{sec:cs}. The relevant part is obtained by going
to transverse position $\B$-space and setting $\B=0$ in the wavefunction
or equivalently one can integrate out the transverse momenta $\Kp$ in
momentum space as shown in \eref{eq:prob_amp}. The other distribution 
amplitudes or wavefunctions that we need is the colour octet of the 
$\c_J$. These have to be constructed because the importance of the 
colour octet in exclusive reactions has not been realized up to now.
For $\c_J$ charmonia with momentum $P$, these can be expressed as 
\be
  |\chi_{\c_J}^{(8)},P \ran\, = \, \frac{t^a_{c\bar c}}{2} \,
                      f_{\c_J}^{(8)}\,\int [\dd z]
                      \f_{\c_J}^{(8)}(z_1,z_2,z_3) \, S_{J\n}^{(8)}(P) 
                      \; |c\bar c g ; P \ran 
\label{coc}
\ee
where $S_{J\n}^{(8)}(P)$ is the spin-wavefunction \cite{bks,wong}. 
It is most convenient to use a product of delta functions for the
distribution amplitude $\f_{\c_J}^{(8)}(z_1,z_2,z_3)$ to set $z = 0.15$
for the constituent gluon and dividing the remaining momentum
fractions $z_1 = z_2 = (1-z)/2$ for the heavy quarks. This value of $z$
was found to be most appropriate in \cite{bks} and the decay
constant $f_{\c_2}^{(8)} =0.90 \times 10^{-3}$ GeV$^2$ was fitted from
$\c_2 \lra \p \p$. The same decay mode is not permitted for $\c_1$ and
the value of $f^{(8)}_{\c_1}$ is therefore unconstrained but it is expected 
to be of the same order. We found a somewhat smaller value at 
$f^{(8)}_{\c_1} =0.225 \times 10^{-3}$ GeV$^2$ for $\c_1$ from the 
decay into $p\bar p$. 

The hard part $T_H$ of the singlet contribution 
is essentially the $\B =\B' =0$ version of \eref{eq:T_H} whereas that
of the octet contribution now comes from an evaluation and summation 
over all members of the ten contributing groups. More explicitly the octet 
hard perturbative part can be written as 

\bea \{T^{J\; (8)}_H (x,y) \}_{\l_1,\l_2,\l_3} 
    &=& i\; (4\p\a_s(m_c))^3 \sqrt{4\p\a^{soft}_s} \; \mc^7 \nonumber \\
    & & \times \sum_{{g \in Groups} \atop {m \in Members}} 
        S_g P_{g\,m}(x,y) \{N^J_{g\,m}(x,y)\}_{\l_1,\l_2,\l_3} \; ,
        \nonumber \\
\eea
where $P_{g\,m}(x,y)$ is the product of propagators of the member $m$
graph of group $g$ and $S_g$ is a symmetry factor of group $g$ to take care of 
similar potential groups that are related to this group by simple change of 
variables. The $\{N^J_{g\,m}(x,y)\}_{\l_1,\l_2,\l_3}$ contains the light 
constituent quark helicity dependence of the outgoing
nucleon and it denotes the numerator of this member $m$ graph. 
The convolution of $T_H$ with the distribution amplitudes can be expressed 
in the form of decay form factors for the octet contributions 
\bea \cb_J^{p \; (8)} 
    &=& \sum_{\l_1,\l_2,\l_3 =\pm} 
        {{f^{(8)}_{\c_J}\;\s_J} \over {2 m_c\; (4 m_c)^4}} 
        \int \dxb \dyb \; \{T^{J\; (8)}_H (x,y) \}_{\l_1,\l_2,\l_3} \;
        \| \hat\j^p(x,0) \hat\j^p(y,0) \|_{\l_1,\l_2,\l_3}  \; ,
        \nonumber \\
\label{eq:o_df}
\eea
where $\s_1 =1/\sqrt{2}$ and $\s_2 = 1$ as given below \eref{eq:s_df},
and the product of proton wavefunction factor 
$\|\hat\j^p \hat\j^p\|_{\l_1,\l_2,\l_3}$ is similar in form to the
wavefunction factor enclosed within braces in \eref{eq:s_df} but
this one has helicity dependence and is evaluated at $\B=\B'=0$. 

The hard part $T^J_H$ of individual feynman diagrams can be readily calculated
using the usual feynman rules and the gauge choice by the same name. 
For example, the numerators and propagator denominators of our example 
member graphs shown in \fref{f:col_neu} (b), \fref{f:ot_g} (a) and (b) for
some chosen helicity combinations are
\bea P_{5,{\srm{\fref{f:col_neu} (b)}}}(x,y) 
     \{N^J_{5,{\srm{\fref{f:col_neu} (b)}}}(x,y)\}_{-++} &=& 
     (-1)^J \frac{\sqrt{2}}{6} 
    \frac{z \{1 -2 x_2 -4 (x_1+x_2-z_1)(y_1-z_1) \}}
         {\{ (z_1-x_1)(z_1-y_1) -1/4 \} \mc^2}                \nonumber \\
    \times \fx & & \frac{1}{\{ (z_2+z-x_3)(z_2+z-y_3) -1/4 \} \mc^2} \;
                                                              \nonumber \\
    \times \fx & & 
    \frac{1}{x_1 y_1 \mc^2} \; \frac{1}{x_2 y_2 \mc^2} 
    \frac{1}{z(x_3-z) \mc^2}                                  \nonumber \\
    \times \fx & &
    \frac{1}{(x_3-z)(y_3-z)\mc^2 +\r^2}  \; ,
\label{eq:g5_n}
\eea
\bea P_{10,{\srm{\fref{f:ot_g} (a)}}}(x,y) 
     \{N^J_{10,{\srm{\fref{f:ot_g} (a)}}}(x,y)\}_{+\pm\mp} &=& 
     \frac{\sqrt{2}\{2(1-z)-(-1)^J (y_2+y_3-z)\}}{(1-x_1-z)(1-y_1-z)\mc^2+\r^2}
                                                             \nonumber \\
     \times \fx & & 
     \frac{1}{(1-z)^2 \mc^2} \; \frac{1}{(1-z)(1-y_1-z) \mc^2}    
                                                             \nonumber \\
     \times \fx & & \frac{1}{x_2 y_2 \mc^2} \;
     \frac{1}{x_3 y_3 \mc^2} \; \frac{1}{\mc^2}                \; ,
\label{eq:g10_n}
\eea
and
\bea P_{4,{\srm{\fref{f:ot_g} (b)}}}(x,y) 
     \{N^J_{4,{\srm{\fref{f:ot_g} (b)}}}(x,y)\}_{+-+} &=& 
     - \frac{5 \sqrt{2}}{3} 
     \frac{(-1)^J (x_2-y_2)}{(1-x_2-z)(1-y_2-z) \mc^2 +\r^2 } \nonumber \\ 
     \times \fx & & \frac{1}{\{ (z_2-x_2)(z_2-y_2) - 1/4 \} \mc^2} 
                    \frac{1}{x_1 y_1 \mc^2}                   \nonumber \\ 
     \times \fx & & \frac{1}{x_2 y_2 \mc^2} \;
     \frac{1}{x_3 y_3 \mc^2} \; \frac{1}{\mc^2}               \; ,
\label{eq:g4_n}
\eea
respectively. To save on typing, we dropped $+i\e$ in the propagators but
its presence should be implicitly understood. The $\r^2$ in some of the 
denominators above does not come with the propagators but has been inserted
by hand. The purpose of this will be explained below. 
The explicit values of $S_g$ and the form of all other $P_{g\,m}(x,y)$, 
$\{N^J_{g\,m}(x,y)\}_{\l_1,\l_2,\l_3}$ and
$\|\hat\j^p \hat\j^p\|_{\l_1,\l_2,\l_3}$ can be found in \cite{wong}. 

The above expressions are merely those of three example graphs. The full 
calculation is tedious but fortunately all necessary algebra can be handled 
by computer. Our treatment and method of dealing with the colour octet
contribution are largely based on and parallel to those in 
\cite{bks,bks2} and the various parameters obtained there are used in the 
present calculations. In particular $\a_s$ is set at the usual scale 
$m_c$ and in any case it is tied to the fitted octet decay constant 
of $\c_J^{(8)}$, and it was assumed $\a_s^{\srm{soft}} =\p$ for this special 
coupling of the constituent gluon to the perturbative part as in \cite{bks}.
This coupling is supposed to be softer than the others and therefore 
requires special treatment. Since the factors
$f^{(8)}_{\c_J} \sqrt{\a_s^{\srm{soft}}}$ always appear together both
in $\c_J \lra \p \p$ and the present $\c_J \lra p\bar p$, the detail of
each individual factor is therefore less important than the combination. 
If the calculation were done within MHSA, both $\a_s^{\srm{soft}}$ and 
its value need not be worried. The latter would be fixed dynamically. 
Once the constituent gluon has been dealt with and put into the 
perturbative hard part, the calculation itself is essentially 
the same as other calculations using SHSA scheme. One calculates $T_H$ 
perturbatively from all contributing graphs and then convolutes it with 
the distribution amplitudes of the hadron involved to get the probability 
amplitude. 

We now return to the $\r^2$ in \etrref{eq:g5_n}{eq:g10_n}{eq:g4_n} above. The
purpose of inserting $\r^2$ is to prevent two poles within one propagator
from occurring simultaneously which is possible in the convolution integral
in \eref{eq:o_df}. When that happens, the $i\e$ prescription cannot
handle such singularities. For the expression in \eref{eq:g5_n} the
problem occurs when $x_3 =y_3 =z$ and for \ederef{eq:g10_n}{eq:g4_n} this is
at $x_1 = y_1 = 1-z$ and $x_2 = y_2 = 1-z$, respectively. The presence
of $\r^2$ removes this kind of singularities in the integral. 
The reason that they are there is because of the collinear approximation 
of SHSA. Within this approximation, the hadron internal transverse 
momentum dependence of the perturbative hard part is dropped 
$T_H(x,y,\Kp) \lra T_H(x,y)$ in favor of hard momenta running through the 
internal lines of the relevant feynman graphs. As a consequence, 
they are dropped from the denominators of the propagators as well 
under the assumption that the large virtualities of the virtual 
propagating particles dwarf any such dependence. Had we relinquished 
our hold of the collinear approximation and calculated within the MHSA,
such singularities simply do not exist. In the MHSA scheme, the
internal transverse momenta of the hadrons involved are kept everywhere
and they therefore serve as a natural cutoff of such singularities.
From this example, the use of $\r^2 = \lan \Kp^2 \ran$, the average
transverse momentum square of the nucleon, as a regulator of these 
singularities naturally suggested itself. Although this treatment is 
somewhat ad hoc within SHSA, it is nevertheless practical and indeed 
has been used in \cite{bks}. 

Next there is the issue of gauge invariance. The common way of expressing
hadron wavefunctions in terms of distribution amplitudes means that 
the energy-momentum of the hadron must be divided into portions and assigned 
to its partonic constituents. In the case of charmonium in a colour octet,
the energy-momentum is therefore assigned as $P=P_c+P_{\bar c}+P_g$ 
with $P_c =z_1 P$, $P_{\bar c}=z_2 P$ and $P_g = z P$. Since $z_1 =z_2$
the mass of the constituent heavy quark is then $z_1 M_{\c_J} \simeq z_1 \mc$
whereas feynman rules do not take into account of particles existing 
in a bounded, confined state and therefore give the standard heavy fermion 
propagator $1/(\g\cdot P_c-m_c)$ with the usual heavy quark mass $m_c$.
Remembering that the probability amplitude of an interaction is 
only gauge invariant, after all feynman graphs of a given
order have been included, provided that the external particles are all on 
their respective mass-shells. The conflict of $z_1 \mc \not= m_c$ thus
becomes a source for the violation of gauge invariance. This only arises in 
the graphs of group (3), (4) and (5). If one checks for gauge invariance, after 
many cancellations, the few remaining gauge violating terms are all proportional 
to $(1-2 z_1) m_c = z m_c$ as expected. Combining with other factors, these
terms first enter at $\co(z^3)$ in the numerators. Our calculations are
therefore gauge invariant up to $\co(z^2)$. As mentioned in \cite{bks2}, one 
could choose to make the results completely gauge invariant by using 
$(1-z)m_c$ in the propagator instead of $m_c$ but because of $z=0.15$, 
these gauge-invariance violating contributions are small and therefore do 
not worry us too much. 

Our results are shown in Table \ref{tab:res}. The colour singlet-octet
combined results are given along with those from the singlet alone. 
The singlet results here differ from those in Table \ref{tab:sing_nucl}
because of the somewhat different values of the octet decay constants 
and the different treatments of the coupling and details between 
the two calculational schemes. But we are satisfied by the fact 
that the results are of the same order of magnitude. For 
experimental errors, the readers can consult ref. \cite{bes} 
and for theoretical ones, they will not be presented at this stage 
since they are tied closely to the experimental ones because the
octet decay constants were obtained in \cite{bks,bks2} by the requirement
of a reasonable agreement with two different sets of experimental measurements.
As mentioned before, the disagreement in the measurements between 
PDG and BES collaboration forces us to settle around the average of the 
two sets of results. This is also the case for the decay mode 
$\c_J \lra \p \p$. Once a better agreement amongst the experiments is 
reached, the decay constants will have to be adjusted accordingly
and estimates of theoretical errors be made. 

As we mentioned already in Sec. \ref{sec:cs}, in the case that the 
true values are close to either the PDG or the BES set. Colour singlet 
alone cannot account for the measured decay widths. We have outlined and 
described our method to include the colour octet. As can be seen from
above, with the colour octet contribution the theoretical situation is 
in a much better shape. Note that the $\c_J$ singlet wavefunction is
well known and in the MHSA, there is no free parameter in
the calculation. As to the colour octet wavefunction, that of
the $\c_2$ is completely fixed by $\c_2 \lra \p \p$ and therefore
there is no remaining freedom left in $\c_2 \lra p \bar p$. 
The value of $f^{(8)}_{\c_1}$ although cannot be fixed by the
pseudo-scalar decay mode like the $\c_2$, the smaller but still at 
the same order value as that of the $\c_2$ can be viewed as 
a consistent result. Since $\c_1$ is a somewhat different P-wave 
charmonium from its even spin partners, one is not too surprising that 
the result is as it is. 

Before summarizing, we would like to address the issue of the validity
of our perturbative calculations. Although the charmonium mass
$\mcj \simeq$ 3.5 GeV is not extremely large, the scales of the coupling
are set essentially by the virtualities of the intermediate gluons and not, 
for example, by the energies of the final partons. For annihilation
into three gluons, each gluon will have about 1.1 GeV and even with a 
large value for $\lqcd =$ 0.25 GeV, the coupling will be about 
$\a_s =$ 0.508. It is not too bad for a perturbative calculation.
On the other hand, if one compares the combined colour singlet and 
octet results in Table \ref{tab:res} with those of the colour singlet 
alone in Table \ref{tab:sing_nucl}, one might be alarmed with the 
apparently large combined results when compared to the smaller 
singlet results and start questioning the validity of the 
perturbative calculations. First, to compare the singlet widths with the
combined singlet and octet widths directly is misleading. A more proper
comparison is between the singlet amplitudes with the combined amplitudes
or equivalently one compares the square root of the widths. 
If this is done, it can be established that the combined amplitudes
are only a factor of 4.5 for $\c_1$ and 3.0 for $\c_2$ larger than 
the corresponding singlet ones, so the octet parts are nowhere 
near completely dominating the results. On the contrary, 
both the singlet and octet parts contribute comparably to the final results. 
Second, with proton-antiproton in the final states and not pions as in
\cite{bks}, one might worry about getting too close to the boundary of
the perturbative region. Fortunately, the calculation of the colour 
singlet contribution in Sec. \ref{sec:cs} and of $J/\J$ decay into 
baryon-antibaryon pair in ref. \cite{bk} both within the MHSA showed 
that the bulk of the contribution came not from large coupling 
non-perturbative region. This should serve as an example of even with 
a pair of baryons in the final states, the calculational scheme of SHSA 
and MHSA are still under perturbative control and can be used 
self-consistently. Third, although it is not possible to do a full calculation
in the MHSA, not all ten groups of Feynman diagrams in the colour octet 
contribution lead to unmanageably high dimensional integrations.
For those that can be calculated numerically with sufficient accuracy,
we have checked that the bulk of the contributions did indeed come
from smaller values of the coupling below $\a_s =$ 0.5. Using a
cutoff of $\a_s^{\srm{cutoff}} =$ 0.5, well over 50 \% of the contributions
were included and with $\a_s^{\srm{cutoff}} =$ 0.6, 98 \% were 
included. Of course, we have not confirmed reasonable widths within
the MHSA due to the complexity of the calculations and the limitation of 
present days computing power, but this should still serve to demonstrate 
to a certain extent that the theoretical arguments presented above are 
reasonably sound and that the calculational schemes used are self-consistent. 

To summarize, we have provided arguments in support of the inclusion
of colour octet in P-wave charmonium decay and by explicit calculation,
have shown that although the same infrared divergence in inclusive 
decay does not exist in exclusive process, for consistency this next higher 
Fock state of the P-wave charmonium must still be included. 
With that, the would-be contradiction discussed at the end of 
Sec. \ref{sec:co_needed} is resolved. The applicability of our arguments 
is obviously not restricted only to charmonia. In fact, they are true 
and should work even better for heavier quarkonia such as $b \bar b$.

\section*{Acknowledgments}

The author would like to thank most of all P. Kroll for introducing him 
to the very interesting subjects of the hard scattering scheme and the colour 
octet picture of quarkonium. The author benefited greatly from his initial 
guidance of this research and from many useful discussions. The author also 
thanks the computational physics group of K. Schilling in Fachbereich 
Theoretische Physik, Universit\"at Wuppertal for using their valuable computer 
resources and for the Nuclear and Particle Physics Section and IASA Institute 
at the University of Athens for kind hospitality where part of this work was 
done. This work was supported by the European Training and Mobility of 
Researchers programme under contract ERB-FMRX-CT96-0008.

\newpage

\section*{Figure Captions}

\begin{itemize}

\item[\fref{f:incl}]{The leading decay process for (a) $\c_0$ and $\c_2$
and (b) $\c_1$. A possible decay of the higher Fock state, the 
colour octet, of the $\c_J$ is shown in (c). In the infrared limit 
of the outgoing gluon, (b) and (c) are degenerate.}

\item[\fref{f:sing}]{Feynman graphs for the colour singlet $\c_J$ decay into
proton-antiproton.}

\item[\fref{f:alpha_crit}]{Percentage of the colour singlet contribution 
to the $\c_J \lra p\bar p$ decay width that comes from
region of values of $\a_s$ below some critical cutoff value 
$\a_s^{\srm{crit.}}$. The solid and dot-dashed line are for $\c_1$
and $\c_2$, respectively. This shows that the bulk of the
contributions in fact comes from $\a_s < 0.6$ region.} 

\item[\fref{f:col_neu}]{Example of two ways to achieve colour neutralization: 
(a) the constituent gluon enters the nucleon as part of the constituent 
in the nucleon after exchanging a gluon with the other constituents, 
(b) this gluon ends in the perturbative hard part $T_H$.}

\item[\fref{f:ot_g}]{Example of $c\bar c$ annihilates through (a) one and (b) 
two gluons. Graph (a) is impossible in the colour singlet contribution.}

\item[\fref{f:fig_basic}]{The feynman graphs that represent the ten basic 
groups from which the graphs of the colour octet contributions are generated.
Each graph is assigned a group number as indicated.}

\end{itemize}

\vspace{1.0cm}

\section*{Table Captions}

\begin{itemize}

\item[\tref{tab:mass_dim}]{The mass dimensions of the decay constants of the 
hadrons discussed in the text. They can be deduced from 
\ederef{eq:prob_amp}{eq:wfun_norm}. }

\item[\tref{tab:qno}]
{The make-up of the parity $\cp$ and C-parity $\cc$ of each colour 
singlet and octet state of the $J/\J$ and $\c_J$ for states with up to three
constituents.}

\item[\tref{tab:sing_nucl}]{Comparing colour singlet contribution of $\c_J$ decay 
into $p\bar p$ to the experimental data.}

\item[\tref{tab:res}]{The colour singlet and octet combined partial decay width
within the SHSA of $\c_J$ decay into proton-antiproton. 
Experimental measurements are also shown for comparison.}

\end{itemize}

\vspace{1.0cm}

\section*{Tables}

\begin{table}[h]
\caption{The mass dimensions of the decay constants of the hadrons
discussed in the text. They can be deduced from 
\ederef{eq:prob_amp}{eq:wfun_norm}.}
\label{tab:mass_dim}
\begin{tabular}{||@{\hspace{.2in}}l||c|c|c@{\hspace{.2in}}||}   \hline
 Hadron         & No. of       & Total Orbital Angular       & Mass dimension      \\
                & Constituents & Momentum $L$  & of $f$ [$f$]   \\ \hline
 $\p$, $K$      &      2       &  0            &    1           \\ 
 $J/\J^{(1)}$   &      2       &  0            &    1           \\ 
 $\c^{(1)}_J$   &      2       &  1            &    2           \\ 
 $p$, $\bar p$  &      3       &  0            &    2           \\ 
 $J/\J^{(8)}$   &      3       &  1            &    3           \\ 
 $\c^{(8)}_J$   &      3       &  0            &    2           \\ \hline
\end{tabular}
\end{table}

\begin{table}
\caption{The make-up of the parity $\cp$ and C-parity $\cc$ of each colour 
singlet and octet state of the $J/\J$ and $\c_J$ for states with up to three
constituents.}
\label{tab:qno}
\begin{tabular}{||@{\hspace{.2in}}l||c|c|c|c|c|c@{\hspace{.2in}}||} \hline
 Hadron            & $L^c_{\bar c}$           & $S_{c\bar c}$  
                   & $L^{c\bar c}_g$          & $N_g$                   
                   & $\cp = (-1)^{N_g+L^c_{\bar c}+L^{c\bar c}_g+1}$ 
                   & $\cc = (-1)^{N_g+S_{c\bar c}+L^c_{\bar c}}$ \\ 
 ($c\bar c$ state) &    &    &             
                   &    &    &                                   \\ \hline
 $J/\J^{(1)}$ ($^3S_1$) & 0 & 1 & 0 & 0 & -1 & -1                \\
 $J/\J^{(8)}$ ($^3P_J$) & 1 & 1 & 0 & 1 & -1 & -1                \\
 $J/\J^{(8)}$ ($^1S_0$) & 0 & 0 & 1 & 1 & -1 & -1                \\ \hline
 $J/\J^{(8)}$ ($^3S_1$)? & 0 & 1 & 0 & 1 & +1? & +1?             \\
 $J/\J^{(8)}$ ($^3S_1$)? & 0 & 1 & 1 & 1 & -1  & +1?             \\ \hline
 $\c_J^{(1)}$ ($^3P_J$) & 1 & 1 & 0 & 0 & +1 & +1                \\
 $\c_J^{(8)}$ ($^3S_1$) & 0 & 1 & 0 & 1 & +1 & +1                \\ \hline
\end{tabular}
\end{table}

\begin{table}
\caption{Comparing colour singlet contribution of $\c_J$ decay 
into $p\bar p$ to the experimental data.}
\label{tab:sing_nucl}
\begin{tabular}{||@{\hspace{.2in}}l|c|c|c@{\hspace{.2in}}||}  \hline
  $J$ & $\Gamma^{(1)}(\c_J\ra p\bar p)$ (eV)         
      & PDG (eV) \cite{pdg} &  BES (eV) \cite{bes}  \\ \hline 
 1    & \ 2.53  & \  75.68   $\pm$ 10.50 & \ 37.84  \\ 
 2    &  16.58  &   200.00   $\pm$ 20.00 &  118.00  \\ \hline
\end{tabular}
\end{table}


\begin{table}
\caption{The colour singlet and octet combined partial decay width
within the SHSA of $\c_J$ decay into proton-antiproton. 
Experimental measurements are also shown for comparison.}
\label{tab:res}
\begin{tabular}{||@{\hspace{.15in}}l|r|r|c|c@{\hspace{.2in}}||} \hline
      & \ \ \ $\Gamma^{(1)}$ (eV)\ \ \ & $\Gamma^{(1)+(8)}$ (eV) 
      & PDG (eV) \cite{pdg} &  BES (eV) \cite{bes}  \\
\hline 
 $\c_1 \lra p\bar p$ &  3.15\ \ \ \ \ &  56.27\ \ \ \ \
                     & \  75.68   $\pm$ 10.50 & \ 37.84  \\ 
 $\c_2 \lra p\bar p$ & 12.29\ \ \ \ \ & 154.19\ \ \ \ \
                     &   200.00   $\pm$ 20.00 &  118.00  \\  \hline
\end{tabular}
\end{table}

\end{document}